# `pgm`: A Python package for free energy calculations within the phonon gas model

Hongjin Wang[1], Jingyi Zhuang[2,3], Zhen Zhang[1], Qi Zhang[1], Renata M. Wentzcovitch[1,2,3]

[1] *Department of Applied Physics and Applied Mathematics, Columbia University, New York, New York, USA, 10027*

[2] *Department of Earth and Environmental Sciences, Columbia University, New York, NY, 10027, USA*

[3] *Lamont–Doherty Earth Observatory, Columbia University, Palisades, NY, 10964, USA*

**Abstract**

The quasi-harmonic approximation (QHA) is a powerful method that uses the volume dependence of non-interacting phonons to compute the free energy of materials at high pressures ($P$) and temperatures ($T$). However, anharmonicity, electronic excitations in metals, or both, introduce an intrinsic $T$-dependence on phonon frequencies, rendering the QHA inadequate. Here we present a Python code, `pgm`, to compute the free energy and thermodynamic property within the phonon gas model (PGM) that uses $T$-dependent phonon quasiparticle frequencies. In this case, the vibrational contribution to the Helmholtz free energy is obtained by integrating the vibrational entropy, which can be readily calculated for a system of phonon quasiparticles. Other thermodynamic properties are then obtained from standard thermodynamic relations. We demonstrate the successful applications of `pgm` to two cases of geophysical significance: cubic $CaSiO_3$-perovskite (cCaPv), a strongly anharmonic insulator and the third most abundant phase of the Earth's lower mantle, and NiAs-type (B8) FeO, a partially covalent-metallic system. This is the oxide endmember of a recently discovered iron-rich $Fe_nO$ alloy phase likely to exit in the Earth's inner core.

Program Summary

*Program Title*: `pgm`

*Developer's repository link:* https://github.com/MineralsCloud/pgm

*Licensing provisions*: GNU General Public License 3 (GPL)

*Programming language*: Python3

*Nature of problem*: The classic quasi-harmonic approximation (QHA) method to compute the vibrational free energy does not apply to physical systems when phonon frequencies have an intrinsic and non-negligible temperature ($T$) dependence. Examples are anharmonic systems or metals with abundant electronic thermal excitations. Both cases introduce an intrinsic $T$-dependence on phonon frequencies.

*Solution method:* The method implemented in `pgm` is based on the phonon gas model where the entropy is well defined for $T$-dependent phonon quasiparticle. The free energy is calculated by integrating the entropy, making it suitable for anharmonic systems or systems with extensive thermal electronic excitations affecting phonon frequencies. The static free energy, the vibrational density of states (VDoS), and the entropy are first computed on sparse $T$- and $P$-grids. The entropy is then suitably interpolated on denser user-specified grids for integration.

*Additional comments, including restrictions and unusual features*: The package allows it to be run directly in the command line. It can also be incorporated into other programs. We implemented Just-in-time (JIT) compiling and parallel computing techniques [1] in `pgm` Python code to speed up the numerical calculation.

# 1. Introduction

The thermodynamic properties of materials at extreme conditions, i.e., high temperatures ($T$) and pressures ($P$), are important in many fields, e.g., chemical engineering, Earth and planetary science, etc. In Earth sciences, thermodynamic properties of minerals improve the understanding of Earth's thermal structure, heat flow, internal dynamics, etc. Experimental measurements of materials properties at planetary interior conditions, e.g., up to ~360 GPa and ~6000 K, are very challenging. Therefore, predictive computational studies are extremely important since they can provide detailed information about these properties. *Ab initio* techniques based on density functional theory (DFT) [2,3] have been shown to be accurate in a broad class of materials under a variety of conditions. *Ab initio* molecular dynamics (MD) [4,5] has been most useful for the study of melts. However, the numerical efficiency of MD is constrained by two factors: ergodicity requires long simulation times and converged thermodynamic properties require large simulation cells. For crystalline solids at certain pressure and temperature conditions (referred to as "*P-T*"), both difficulties are successfully resolved by the use of the quasiharmonic approximation (QHA) [6,7]. In the framework of the QHA, the vibrational free energy, $F_{vib}$, at high *P-T*s is computed using the volume ($V$) dependence of non-interacting phonon frequencies. $F_{vib}$ consists of the total energy of a collection of non-interacting harmonic oscillators at a well-defined $V$ and $T$. Therefore, the QHA also accounts for nuclear quantum (zero-point) effects and is the appropriate method to use at low $T$s. In most cases, the QHA is appropriate for harmonic or weakly anharmonic solids up to ~ 2/3 of the melting temperature [6,8,9]). The QHA is computationally less demanding than MD but still requires intensive computations to obtain the vibrational density of states (VDoS) for each optimized volume. However, anharmonicity caused by phonon-phonon interactions (e.g., [10]), thermal electronic excitations in metals [11], or both (e.g. [12]), introduce an intrinsic *T*-dependence on phonon frequencies, rendering the QHA inadequate.

In this paper, we present the code name `pgm`, a Python implementation of a free energy calculation scheme that can be used with *T*-dependent phonon frequencies. Inspired by the phonon gas model (PGM), the code calculates the free energy by integrating the entropy, which is expressed in terms of *T*-dependent phonon quasiparticle frequencies. The code is capable of calculating the thermal equations of state and a variety of thermodynamic properties. It requires only the static free energies, electronic entropy (if applicable) and VDoSs calculated on a sparse temperature-volume (*T-V*) grid. The users can specify a denser *T-V* range to obtain continuous thermodynamic properties. The thermodynamic properties of interest can be selected using the configuration file of the code. The input VDoSs can either be harmonic phonons without considering the effects of phonon-phonon interactions, or anharmonic phonons, including the effects of phonon-phonon interactions, depending on the nature of the studied system. Harmonic phonon frequencies can be obtained by either density functional perturbation theory (DFPT) [13] or the finite displacement method [14,15]. Anharmonic/renormalized phonon frequencies can be obtained by methods such as the phonon quasiparticle approach [16,17], temperature-dependent effective potential [18], self-consistent phonon [19], self-consistent *ab initio* lattice dynamics [20], and stochastic self-consistent harmonic approximation [21], etc.

`pgm` has a wide range of applications in the study of complex systems. We present two different systems in the Earth's interior to demonstrate the applicability of `pgm`: cubic $CaSiO_3$-perovskite (cCaPv), and B8-FeO. cCaPv is a major mineral phase of the Earth's lower mantle which is a strongly anharmonic system [22], exhibiting phonon instabilities at low temperatures [23]. B8-FeO is a vital alloying component in the Earth's core [24,25]. It is a partially covalent-metallic oxide endmember of a recently discovered iron-rich $Fe_nO$ alloy [24–27] phase likely to exit in the Earth's inner core. In both complex systems, `pgm` analysis can provide insightful results for understanding their thermodynamic properties.

This paper is organized as follows. The next section reviews the formalism of the phonon gas model. Section 3 describes the code, its installation, and execution. In Sections 4 and 5, we show the structure of the input and output files for the program, respectively. In Section 6, we present the results of two examples calculated by the `pgm` code. Our conclusions are summarized in section 7.

## 2. Methods used in the pgm code

### 2.1 Phonon gas model (PGM)

Anharmonicity, thermal electronic excitations in metals, or both, introduce an intrinsic *T*-dependence on the phonon frequencies, which makes the QHA free energy formula inadequate [12,28,29]. Henceforward, we use a free energy calculation scheme based on the phonon gas model (PGM) that uses as input *T*-dependent phonon quasiparticle VDoSs [10,16,23,28,30].

In the PGM, phonons or phonon quasiparticles are non-interacting. The vibrational entropy in this system is the summation of quasiparticle mode contributions over the Brillouin zone,

$$S_{vib}(T,V) = k_B \sum_{q,s} \left((1 + n_{qs})\ln(1 + n_{qs}) - n_{qs} \ln n_{qs}\right), \qquad (1)$$

where $n_{qs}(T,V) = 1/\left[\exp\frac{\hbar \widetilde{\omega}_{qs}(T,V)}{k_B T} - 1\right]$ is the Bose-Einstein distribution with $T$-dependent frequencies with wave-vector $\boldsymbol{q}$ in branch $s$. $\hbar$ and $k_B$ are the Planck and Boltzmann constants, respectively.

There are generally two calculation types, corresponding to systems with and without an approachable ground state, respectively (see flowchart). In systems with an approachable ground state, entropy is integrated from zero temperature, $T = 0$. In this case, with $S_{vib}(T)$ obtained, the Helmholtz free energy at constant volume $V$ is calculated as,

$$F(T) = E_0 + \frac{1}{2}\sum_{q,s} \hbar \omega_{qs}^{0K} - \int_0^T S(T')dT'. \qquad (2)$$

$E_0$ is the DFT total energy obtained at $T = 0$ K. $\omega_{qs}^{0K}$ is harmonic phonon frequency obtained at $T = 0$ K and $\frac{1}{2}\sum_{q,s} \hbar \omega_{qs}^{0K}$ is the zero-point energy $E_{zp}(V)$. The total entropy $S = S_{vib}$ for insulators and $S = S_{el} + S_{vib}$ for metals. $S_{el}(T)$ is the AIMD run average of the electronic entropy obtained with the Mermin functional. [31,32] The temperature-dependent frequencies $\widetilde{\omega}_{qs}$ used in $S_{vib}$ calculations can be: 1) $T_{el}$-dependent harmonic phonon frequencies due to electronic excitations for metals [28,33], 2) $T$-dependent phonon quasiparticle frequencies due to phonon-phonon interactions, i.e., anharmonicity for anharmonic insulators [16,23,34], and 3) $T$-dependent phonon quasiparticle frequencies due to both phonon-phonon interactions and electronic excitations for anharmonic metals, where $T_{el} = T$ [12,29,35].

In systems without an approachable ground state and the structure is dynamically stabilized by anharmonicity only at finite temperatures, entropy can be integrated from a finite temperature, $T_0$, at which both the structure and phonon quasiparticles are stable [12,23]. In this case, with $S_{vib}(T)$ obtained, the Helmholtz free energy at constant volume $V$ is calculated as,

$$F(T) = E(T_0) - T_0 S(T_0) - \int_{T_0}^T S(T')dT'. \qquad (3)$$

$E(T_0)$ is the time-averaged internal energy obtained from *ab initio* molecular dynamics (AIMD) simulations at $T_0$. $S$ is obtained similarly to that in the first case. In this second case, which is handled by AIMD at classical regime, the zero-point energy is not included.

The pgm code offers a highly convenient solution for researchers studying different types of systems, by equally reading in temperature-dependent frequencies without more complexity. The user has the flexibility to determine which effects they wish to consider in their study of the system: electronic excitations for metals, phonon-phonon interactions for anharmonic insulators, or electronic excitations plus phonon-phonon interactions for anharmonic metals, etc. The pgm code greatly expands the scope of materials' free energy and thermodynamics studies beyond the quasiharmonic approximation codes that limited to dealing only with temperature-invariant phonon frequencies.

## 2.2 Equation of State fitting

The calculation of the free energy and the derivation of other thermodynamic properties require a fine *T-V* grid. The interpolation on an initial discrete *T-V* mesh is done separately and sequentially for *T* and *V*.

The *T* interpolation at the coarse volume grid is particularly delicate as *T* enters in two different ways depending on whether the system is an insulator or a metal. For an insulator, $T_{el} = 0$ always holds, and only $T_{ion}$ is relevant, $T = T_{ion}$, as applied by the phonon quasiparticle spectrum calculations [33]. For a metal, *T* enters as $T_{el}$ in the Mermin functional, $T = T_{el}$. The *T* equations always hold, and in practice, one calculates $F(T_{el}, T_{ion}, V)$ and subsequently makes $T = T_{el} = T_{ion}$ [12,28]. In both cases, the quasiparticle spectrum $\widetilde{\omega}_{qs}(T)$ is initially provided at 5-10 temperatures, and interpolated using quadratic form of polynomials, which gives a good fit to $\widetilde{\omega}_{qs}$ This allows to obtain $S_{vib}(T)$ on a fine *T*-mesh, which is essential to compute the integral along temperatures. For metals, we also need to calculate $S_{el}(T)$ on the same fine *T*-grid, which is available as an MD ensemble average in the coarse *T*-grid, the same quadratic form of polynomials as for $\widetilde{\omega}_{qs}(T)$.

The *V* interpolation is also necessary, and is done initially on the sparse input volumes (usually in the size of 6 to 10) and then interpolated on a denser volume grid using the 3rd-order Eulerian finite strain equation of state [36].

By using methods mentioned above, we can accurately calculate thermal properties from sparse input volumes and achieve a more complete understanding of materials properties.

**2.3 Thermodynamic properties**

Once the free energy has been calculated (as described by Eqs. (2,3)) on the designated fine temperature and volume grids, a variety of thermodynamic properties can be computed based on the thermodynamic relationships detailed in Table 1. These properties can be computed numerically with high accuracy.

**3. Description of the program**

**3.1 Flowchart of the `pgm` code**

`pgm` works with Python 3. A comprehensive flowchart depicting the calculation process is available in Figure 1. The procedure is concisely outlined below:

1. Read the data and a user-specified configuration file. A thorough explanation of the input file's format and parameters can be found in Section 4.
2. Calculate the free energy based on the provided data files using entropy integration, using the formalism in Section 2.
3. Determine which thermodynamics properties to compute based on the users' specifications. The results can be calculated on the dense *(T, P)* and *(T, V)* grids.
4. Save all the calculated properties to plain text files, containing a comma-separated value (CSV) table. The names of the output files and the units of the output properties can be found in Table 5.

A description of the core modules used by `pgm` is shown in Table 2.

**3.2 Installation**

We recommend using the Python package manager `pip` (https://pypi.org/project/pip/) to install the `pgm` package. First, install a series of required libraries by executing "`pip install -r requirements.txt`" at the directory that contains `requirements.txt`. Then, change the current directory to the `pgm` top-level directory and run the command "`pip install .`".

Specific installation instructions can also be found in the `README.md` file inside the `pgm` package, and in the page about installation in the `pgm` online documentation in Section 3.4.

**3.3 Program Execution**

To execute the program, the user should first prepare the input files, as outlined in Section 4. Once complete, the user can navigate to the directory containing the `YAML` settings file (hereafter referred to as `settings.yaml`) and execute "`pgm run settings.yaml`" to initiate the calculation process. Upon completion, the thermodynamic property outputs will be saved to the path designated by the user in the settings file.

### 3.4 Examples and documentation

Two examples are given in the `examples` directory to demonstrate the features of `pgm` code, including the cubic $CaSiO_3$-perovskite (cCaPv) example, a strong anharmonic system, and the B8-FeO example, a partial covalent/metallic system at high temperature where electronic excitations cannot be ignored. Both examples give input data files and configurational files.

`pgm` comes with comprehensive documentation that is generated by a documentation generator `Sphinx` (http://www.sphinx-doc.org). To access the documentation, visit online https://mineralscloud.github.io/pgm.

### 4. Input files

The `pgm` code has two types of input files: a user-specified control file, and a series of data files (see Table 3). We use the same file structure used in the Python package `qha` [37] to store phonon calculation information to optimize compatibility. We use `YAML` syntax for the user-specified control file (see http://yaml.org/ for a detailed syntax explanation). All allowed options for the configuration file are listed in Table 4.

### 5. Output files

The output files are a series of plain text files, containing a comma-separated value (CSV) table of the thermodynamics properties specified by the users in the control. The output file contains at least the free energy results without specifying more desired thermodynamic properties. The available output files are listed in Table 5.

### 6. Examples

Here we show the performance of the `pgm` package on two important materials of importance to geophysics and planetary modeling: cubic $CaSiO_3$ perovskite (cCaPv), a highly anharmonic insulator without an approachable ground state, and B8 FeO, a metallic system with an approachable ground state.

### 6.1 Cubic calcium silicate perovskite CaSiO₃ (cCaPv)

CaSiO₃ perovskite (CaPv) is the third most abundant phase in the lower mantle. Among a variety of CaPv phases, the cubic phase with Pm$\bar{3}$m space group is the [38,39] stable form under lower mantle P-T conditions, i.e., 23 < P < 135 GPa and 2000 < T < 4000 K [40,41]. It is a strongly anharmonic phase, and one must obtain the T-dependent phonon quasiparticle spectrum to compute the free energy in Eqs. (2,3). The use of the Mermin functional is not needed since this mineral is an insulator. The *ab initio* calculation reported in this example was carried out in Ref. [23].

Cubic CaPv (cCaPv) is unstable at zero Kelvin or low temperatures, so it does not have an approachable ground state. Accordingly, its harmonic phonon dispersion displays imaginary phonon branches. It is dynamically stable only at high temperatures, e.g., at lower mantle conditions. To overcome such difficulty and calculate cCaPv's thermodynamics, we use the phonon quasiparticle approach [16,17] and the `phq` code [33]. The phonon quasiparticle approach delivers finite-temperature anharmonic phonon quasiparticle dispersion of cCaPv free of imaginary frequencies, which can be used to calculate the thermodynamics within the PGM. The quasiparticle frequencies are extracted from *ab initio* mode-projected velocity autocorrelation functions, which require *ab initio* MD simulations and *ab initio* harmonic phonon calculations. In this example, *ab initio* MD simulations were conducted using VASP [42] on 2 × 2 × 2 supercells (40 atoms). We employed the LDA [3] and the projector-augmented wave method [43] with an associated plane-wave basis set energy cutoff of 550 eV. MD runs were carried out at five volumes (see Figure 4) and six temperatures ranging from 1500 K to 4000 K controlled by the Nosé thermostat [44] covering lower mantle conditions. Each MD ran for 60 ps with a time step of 1 fs. *Ab initio* harmonic phonons were calculated using DFPT [13]. With phonon quasiparticle frequencies obtained, the phonon quasiparticle dispersion is obtained via Fourier interpolation.

Figure 2 shows the comparison of harmonic phonon dispersion at 0 K and anharmonic phonon dispersion at 1000 K. The harmonic phonon dispersion presents unstable phonon modes, which are stabilized by anharmonic interactions only at high temperatures. The stable anharmonic phonon dispersions enable us to apply the `pgm` package to calculate free energy for cCaPv at high-temperature conditions. Apart from the anharmonic phonon spectra, internal energies output by the *ab initio* MD simulations are also required for the thermodynamics calculations. The internal energy is obtained as the ensemble average of the potential energy surface plus the ionic kinetic energy. An additional correction to the internal energy was applied, which brought the computed EoS into full agreement with experimental high-temperature EoSs (see Figure 3) [23]. The thermodynamics of cCaPv were calculated on a *T, P* grid with $dP = 1$ GPa and $dT = 10\ K$.

The free energy and entropy of CaSiO$_3$ are shown in Figure 4 on a (*T, V*) grid. Figure 5 shows thermodynamic properties calculated on the (*T, P*) grid. We compare our results with a previous study, which used Mie–Grüneisen–Debye (MGD) method along the temperature grid [45]. The obtained thermodynamic properties agree very well with the results reported in Ref. [23]. The divergences in the properties calculated by the PGM method and the MGD method indicate the intrinsic anharmonicity in cCaPv phonon dispersions should play an important role in the calculations. The computed bulk modulus $K_s$ varies in a smaller range than the results from MGD calculations, showing that future predictions of seismic wave velocities for this material must incorporate anharmonic effects. The present methodology is important for exploring thermodynamic and thermoelastic properties and phase boundaries in strongly anharmonic systems at high pressures and temperatures.

**6.2 NiAs-type ferrous oxide (B8 FeO)**

FeO is one of the major components of Earth and other terrestrial planets. The NiAs-type structure (B8) of FeO was observed at ground state and up to ~3800 K and ~240 GPa [46], with partially covalent/metallic state [26,27]. Due to the metallic nature of B8 FeO at high pressure and temperature conditions of the Earth's interior, the effects of electronic excitations on the vibrational and electronic entropies need to be taken into account for accurate thermodynamics calculations. Such effects have been demonstrated to be nonnegligible in another metallic system, Fe, in our previous study [28].

Unlike the previous cCaPv example, B8 FeO has an approachable ground state, i.e., the structure and harmonic phonons are stable at 0 K. To include the effects of electronic excitations, the electronic-temperature-dependent harmonic phonon dispersions and electronic entropies were used in the free energy calculations. On top of the effects of electronic excitations considered in this study, phonon-phonon interactions, i.e., anharmonicity can also be included by replacing the temperature-dependent phonon frequencies with phonon quasiparticle frequencies [47] without more complexity.

In this study, the harmonic phonons were calculated using density-functional perturbation theory (DFPT) [13] on 4 × 4 × 2 q-meshes for B8 FeO. DFT and DFPT calculations were performed using the PAW method [43] as implemented in Quantum ESPRESSO [48]. We used the Perdew-Burke-Ernzerhof (PBE) [49] generalized gradient approximation (GGA) to compute the exchange-correlation energy. Electronic excitations at finite temperatures were taken into account by the Mermin functional [31] with Fermi-Dirac smearing. Electronic ground state at 0 K was handled by the Methfessel-Paxton smearing with a width of 0.2 eV [15,29,50].

Figure 6 shows the phonon dispersions of B8 FeO for four different $T_{el}$. All phonon dispersions are calculated at the same volume, but slightly variable cell shape since the latter depends on $T_{el}$. These phonon dispersions do not differ drastically, suggesting a potentially weak dependence of the vibrational free energy on $T_{el}$. Nevertheless, the use of $T$-dependent phonon frequencies in the $S_{vib}$ calculation improves the predictive power of these calculations. Figure 7 shows compares calculated thermal equations of state with available experimental data Ref. [51,52]. We use same color for experimental results and the pgm results to indicate the same temperature. Despite the extreme and challenging conditions of the measurements, we have good agreement with the experimental data. Figure 8 displays variations in the Helmholtz free energy and total entropy on a ($T$, $V$) grid while Figure 9 reports thermodynamic properties on the ($T$, $P$) grid at four different pressures (240, 280, 320, 360 GPa). The solid lines depict results obtained using the pgm code with full thermal electronic excitation effects included. In this case, the free energy is computed using Eq.(2), which is also written as $F(V,T) = E_0(V) + E_{zp}(V) - \int_0^T S(V,T')dT' = F_{mermin}(V,T) + F_{vib}(\widetilde{\omega}(V,T),T)$. In contrast, the dotted lines correspond to results excluding thermal electronic excitations, i.e., $T_{el} = 0$ K and temperature independent phonons. These results are identical to results predicted by the standard quasiharmonic approximation using the qha code [37], where the free energy is obtained using $F(V,T) = E(V) + F_{vib}(\omega(V),T)$. Overall, the results suggest that FeO exhibits complex thermodynamic behavior at high pressures and temperatures, i.e., with a rich interplay of electronic excitation effects specially on the expansivity coefficient ($\alpha$) the property most sensitive to the temperature dependent of phonon frequencies, usually produced by phonon-phonon interactions (intrinsic anharmonicity).

## 7. Conclusions

The `pgm` package can calculate the anharmonic free energy of solids in the thermodynamic limit ($N \rightarrow \infty$) using $T$-dependent phonon frequencies. The implemented method is applicable to weakly or strongly anharmonic insulators or metals up to extreme conditions near melting. As a standard python package, `pgm` can be further integrated into other Python packages or be called in a variety of scripts.

## Acknowledgments

This work was supported in part by the US Department of Energy award DESC0019759 and in part by the National Science Foundation award EAR-1918126 (R.M.W.). This work used the Extreme Science and Engineering Discovery Environment (XSEDE), USA, which was supported by the National Science Foundation, USA Grant Number ACI-1548562. Computations were performed on Stampede2, the flagship supercomputer at the Texas Advanced Computing Center (TACC), the University of Texas at Austin generously funded by the National Science Foundation (NSF) award ACI-1134872.

Tables

**Table 1**: Equations used to calculate thermodynamic properties.

| Property name | Equation |
| --- | --- |
| Pressure $P$ | $P = -\left(\frac{\partial F}{\partial V}\right)_T$ |
| Entropy $S$ | $S = -\left(\frac{\partial F}{\partial T}\right)_V$ |
| Internal energy $U$ | $U = F + TS$ |
| Enthalpy $H$ | $H = U + PV$ |
| Gibbs free energy $G$ | $G = F + PV$ |
| Thermal expansion coefficient $\alpha$ | $\alpha = \frac{1}{V}\left(\frac{\partial V}{\partial T}\right)_P$ |
| Isothermal bulk modulus $B_T$ | $B_T = -V\left(\frac{\partial P}{\partial V}\right)_T$ |
| Grüneisen parameter $\gamma$ | $\gamma = \frac{B_T \alpha V}{C_V}$ |
| Adiabatic bulk modulus $B_S$ | $B_S = B_T(1 + \gamma \alpha T)$ |
| Volumetric heat capacity $C_V$ | $C_V = \left(\frac{\partial U}{\partial T}\right)_V$ |
| Isobaric heat capacity $C_P$ | $C_P = C_V(1 + \gamma \alpha T)$ |

**Table 2**: A description of the core modules.

| Module name | Description |
| --- | --- |
| `settings.py` | Control the calculations processed in this program; provide a set of default calculation settings; read user-specified settings from a `YAML` file. |
| `read_input.py` | Read in the series of volumes and static total energies, and frequencies from *ab initio* calculation. |
| `calculator.py` | Calculate Helmholtz free energy using `pgm` by calculating the vibrational entropy, integrating along the temperatures, and adding the static part. |
| `fitting.py` | Implement finite strain equation of state (EoS) fitting functions. |
| `grid_interpolation.py` | Perform finite strain EoS fitting on the free energies and volumes, and interpolate the fitted function on the automatically generated denser volume grid. |
| `unit_conversion.py` | Provide conversion functions among common units. |
| `thermo.py` | Calculate all the thermodynamic properties listed in the Feature section as well as in Table.1. |
| `v2p.py` | Transform the thermodynamic properties from $(T,V)$ grid to $(T,P)$ grid. |

**Table 3**: Files needed for using the code.

| File name | Description |
| --- | --- |
| `settings.yaml` | Settings file for `pgm` calculation, the user can set the temperatures, pressures, input file structures, and output files. |
| `input-0K`, `input-300K`, `input-1000K`... | Each input data file contains the volumes, static total energies, and frequencies from *ab initio* calculations at its temperature as specified in the input file name. |

**Table 4**: Parameters in the `settings.yaml` file in YAML syntax.

| Parameters | Description |
| --- | --- |
| `NV` | Number of volumes (same as pressures) on the grid |
| `NT` | Number of temperatures on the grid |
| `folder` | The folder or data file name containing the placeholder for the temperatures, i.e., '/%sK/input.txt', './%sK.txt' |
| `initP` | The desired initial pressure in the unit of GPa for calculations |
| `finalP` | The desired final pressure in the unit of GPa for calculations |
| `temperature` | The list of temperatures of the inputs in the unit of Kelvin which are put into the placeholder specified in the folder parameter. |
| `output_directory` | The path to save the output data. The default path is './results/'. |
| `pressure` | Boolean type value. Determine whether to output pressure vs. temperature and volume results |
| `entropy` | Boolean type value. Determine whether to output entropy vs. temperature and volume results |
| `internal_energy` | Boolean type value. Determine whether to output internal energy vs. temperature and pressure results |
| `enthalpy` | Boolean type value. Determine whether to output enthalpy vs. temperature and pressure results |
| `gibbs_free_energy` | Boolean type value. Determine whether to output Gibbs free energy vs. temperature and pressure results |
| `thermal_expansion_coefficient` | Boolean type value. Determine whether to output thermal expansion coefficient vs. temperature and pressure results |
| `isothermal_bulk_modulus` | Boolean type value. Determine whether to output isothermal bulk modulus vs. temperature and pressure results |
| `gruneisen_parameter` | Boolean type value. Determine whether to output the Grüneisen parameter vs. temperature and pressure results |
| `adiabatic_bulk_modulus` | Boolean type value. Determine whether to output adiabatic bulk modulus vs. temperature and pressure results |
| `volumetric_heat_capacity` | Boolean type value. Determine whether to output volumetric heat capacity vs. temperature and pressure results |
| `isobaric_heat_capacity` | Boolean type value. Determine whether to output isobaric heat capacity vs. temperature and pressure results |

**Table 5**: A list of the important output properties as functions of temperature and pressure or volume.

| Property name | Output | Description |
| --- | --- | --- |
| **Free Energy $F$** | `ftv_ev_a3` | Helmholtz free energy (T, V) result. Free energy in unit of ev, volumes in ang3 |
| **Pressure $P$** | `ptv_gpa_K_a3` | Pressure (T, V) result. Pressure in the unit of GPa, temperature in Kelvin, volumes in ang3. |
| **Entropy $S$** | `stv_ev_K_a3` | Entropy (T, V) result. entropy in unit of ev/k, temperature in Kelvin, volumes in ang3. |
| **Internal energy $U$** | `utp_ev_K_gpa` | Internal energy (T, P) result. Internal energy in the unit of eV, temperature in Kelvin, pressure in GPa. |
| **Enthalpy $H$** | `htp_ev_K_gpa` | Enthalpy (T, P) result. Enthalpy in the unit of eV, temperature in Kelvin, pressure in GPa. |
| **Gibbs free energy $G$** | `gtp_ev_K_gpa` | Gibbs Free Energy (T, P) result. Gibbs Free Energy in the unit of eV, temperature in Kelvin, pressure in GPa. |
| **Thermal expansion coefficient $\alpha$** | `alpha_K_gpa` | Thermal expansion coefficient (T, P) result. Thermal expansion coefficient in the unit of 1/K, temperature in Kelvin, pressure in GPa. |
| **Isothermal bulk modulus $B_T$** | `bt_gpa_K_gpa` | Isothermal bulk modulus (T, P) result. Isothermal bulk modulus in the unit of GPa, temperature in Kelvin, pressure in GPa. |
| **Grüneisen parameter $\gamma$** | `gamma_K_gpa` | Grüneisen parameter (T, P) result. Temperature in the unit of Kelvin, pressure in GPa. |
| **Adiabatic bulk modulus $B_S$** | `bs_gpa_K_gpa` | Adiabatic bulk modulus (T, P) result. Adiabatic bulk modulus in the unit of GPa, temperature in the unit of Kelvin, pressure in GPa. |
| **Volumetric heat capacity $C_V$** | `cv_jmol_K_gpa` | Volumetric heat capacity (T, P) result. Volumetric heat capacity in the unit of J·mol/K, temperature in the unit of Kelvin, pressure in GPa. |
| **Isobaric heat capacity $C_P$** | `cp_jmol_K_gpa` | Isobaric heat capacity (T, P) result. Isobaric heat capacity in the unit of J·mol/K, temperature in the unit of Kelvin, pressure in GPa. |

Figures

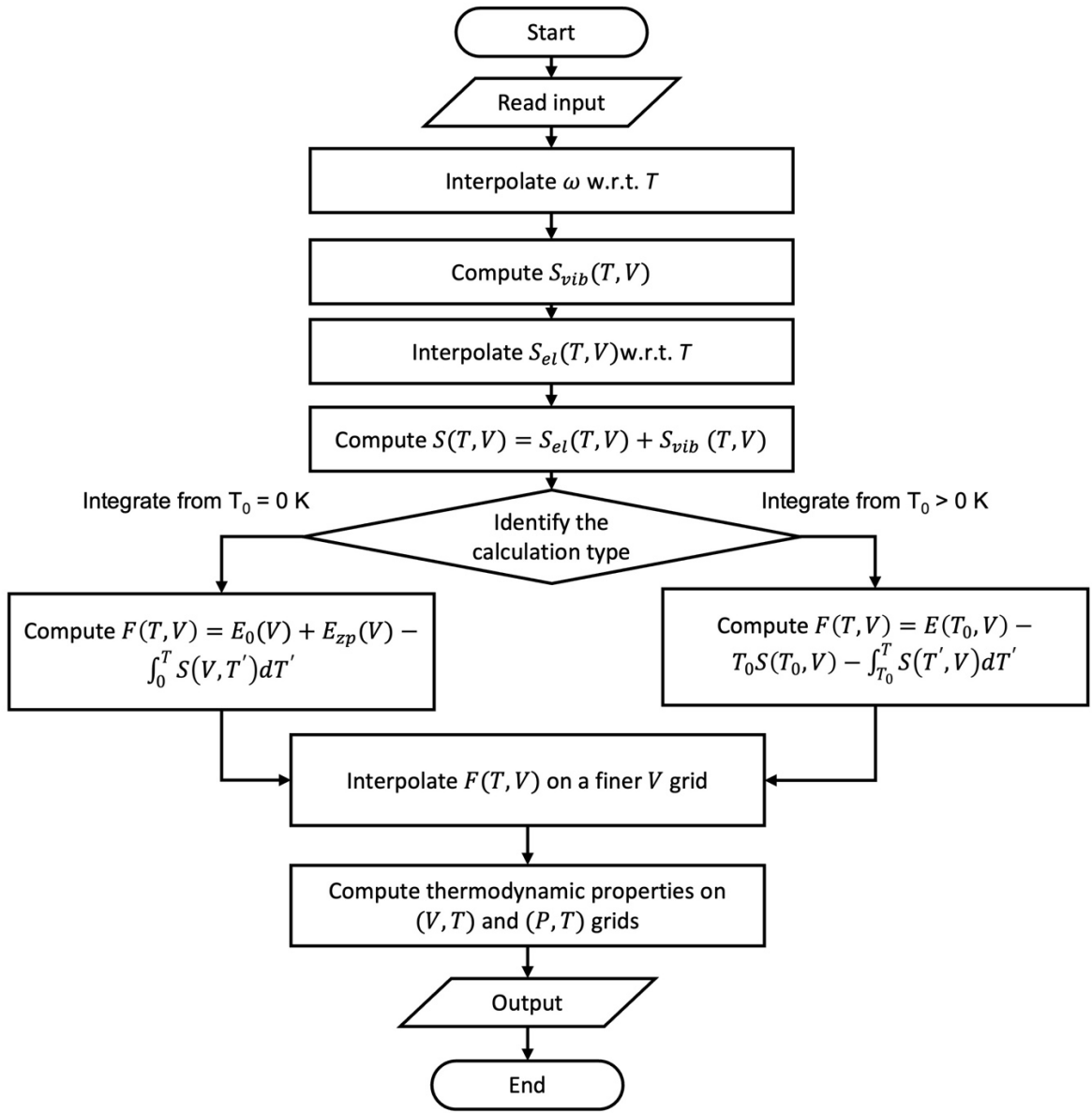

**Figure 1**. Flow chart of the calculations processed in the `pgm` program.

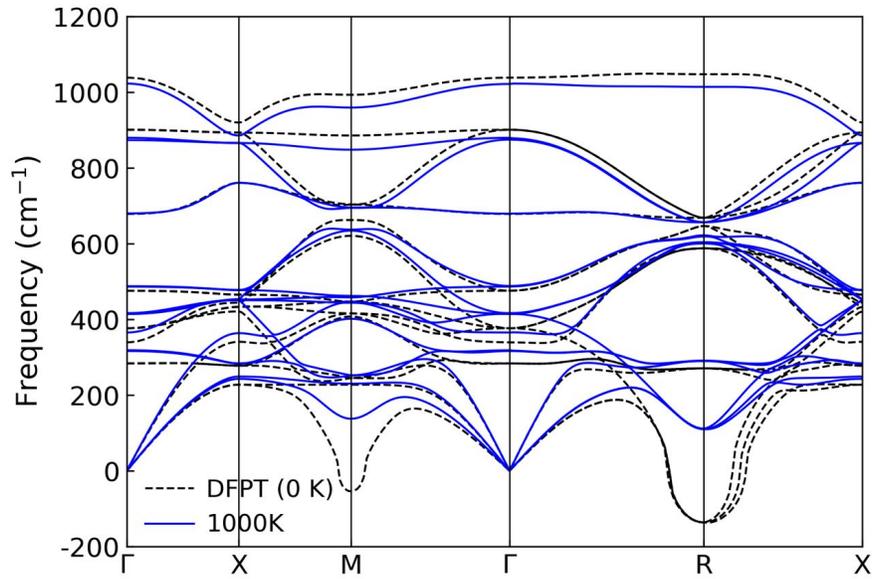

**Figure 2**. Unstable phonons (dashed curves) of cCaPv stabilize at high T (solid blue curves).

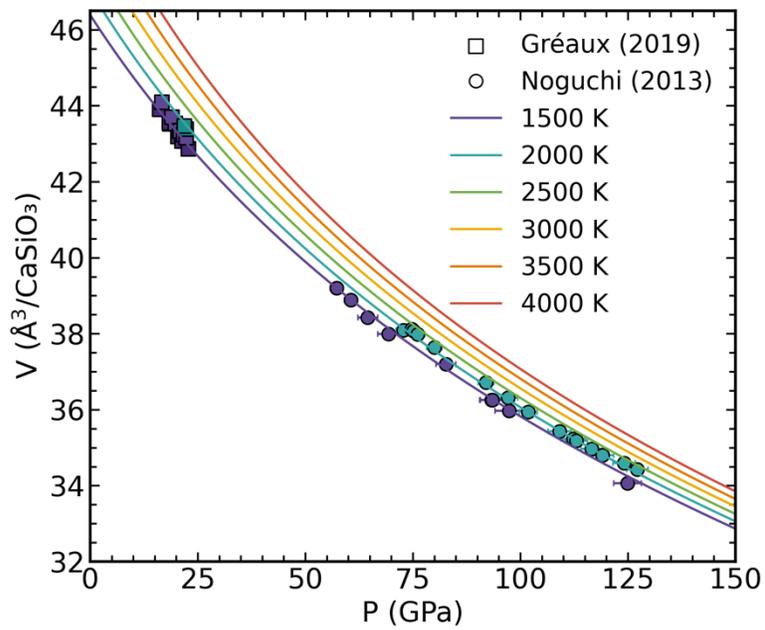

**Figure 3**. Equation of states of cCaPv at various temperatures with comparison to experiments [53,54].

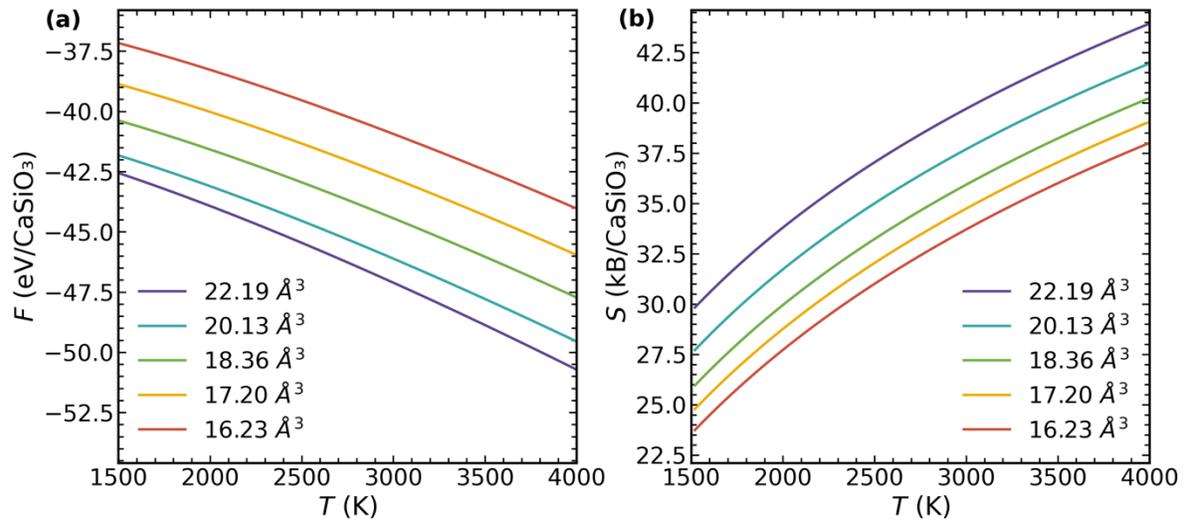

**Figure 4**. (a) Helmholtz free energy $F$ and (b) Entropy $S$ of cubic $CaSiO_3$ perovskite vs. temperatures at a series of volumes (22.19, 20.13, 18.36, 17.29, and 16.23 Å³).

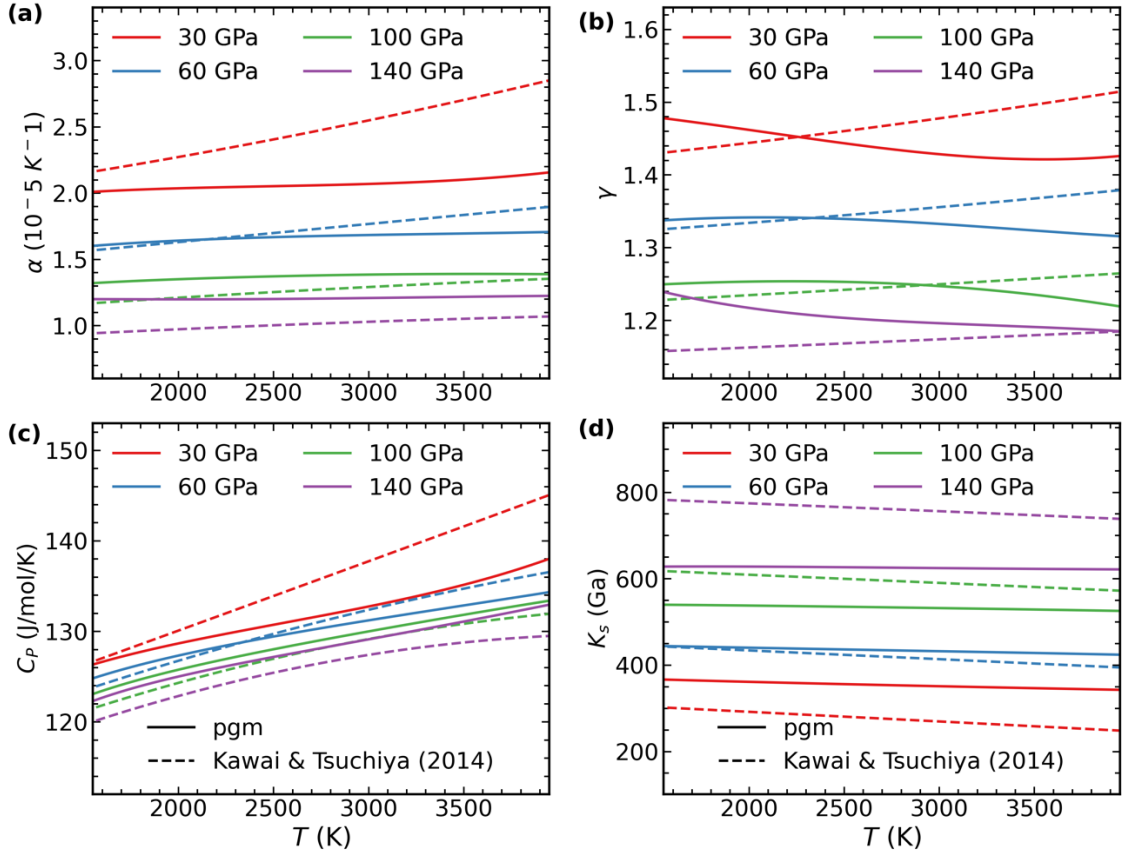

**Figure 5**. Thermodynamic properties of cubic CaSiO$_3$ perovskite (a) thermal expansivity $\alpha$ (b) güneisen parameters $\gamma$ (c) isobaric heat capacity C$_P$ (d) adiabatic bulk modulus K$_S$ vs. temperatures at four different pressures (30, 60, 100, 140 GPa). Solid lines are obtained from the `examples/casio3` using the `pgm` code; dashed lines are results from the previous computational study [45].

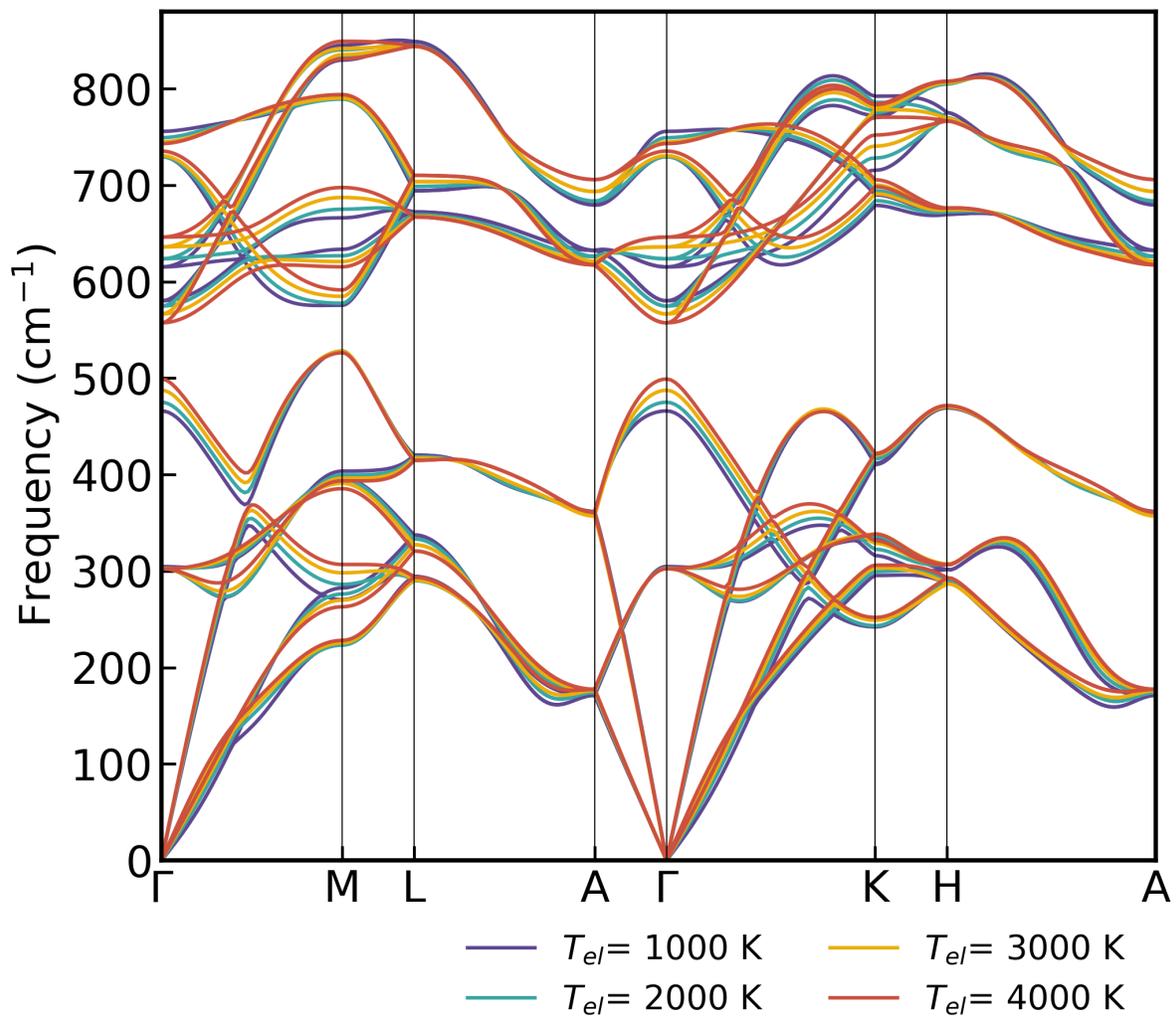

**Figure 6**. Phonon dispersion of B8 FeO calculated at different electronic temperatures (1000K, 2000K, 3000K, 4000K).

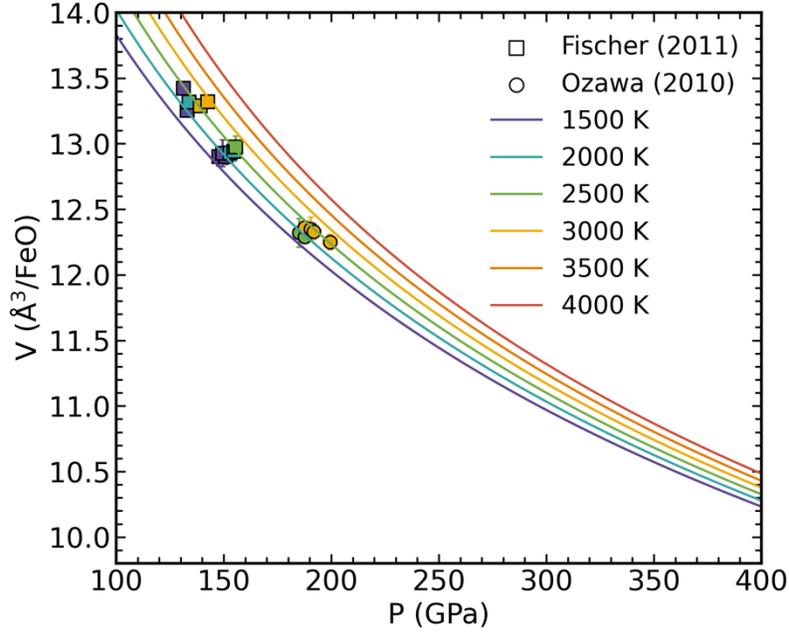

**Figure 7**. Equation of states of FeO at various temperatures with comparison to experiments [51,52].

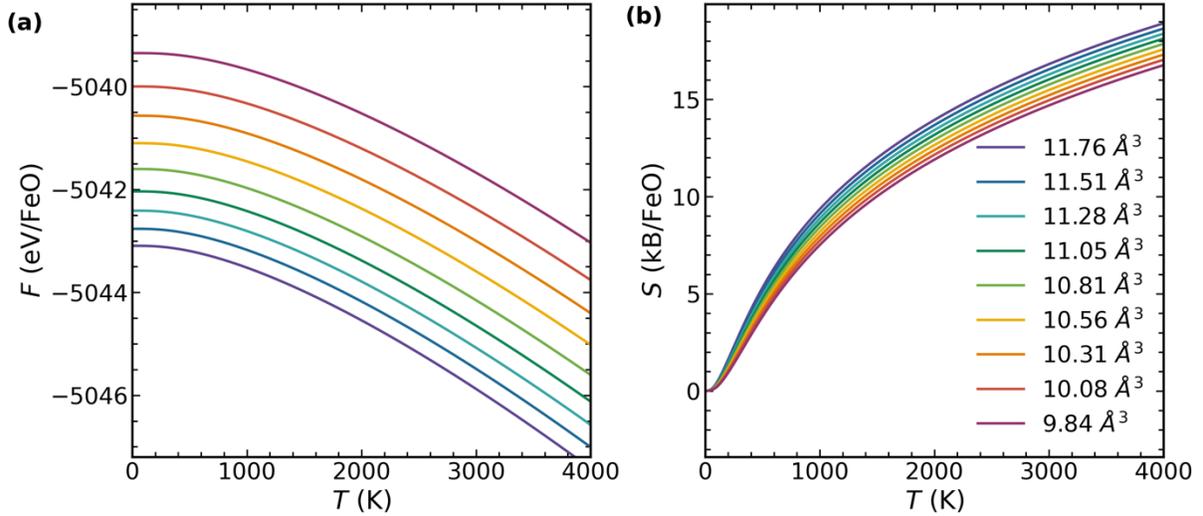

**Figure 8**. (a) Helmholtz free energy $F$ and (b) Entropy $S$ of B8 FeO vs. temperatures at a series of volumes (11.77, 11.54, 11.28, 11.03, 10.79, 10.56, 10.33 10.09, and 9.83 Å$^3$).

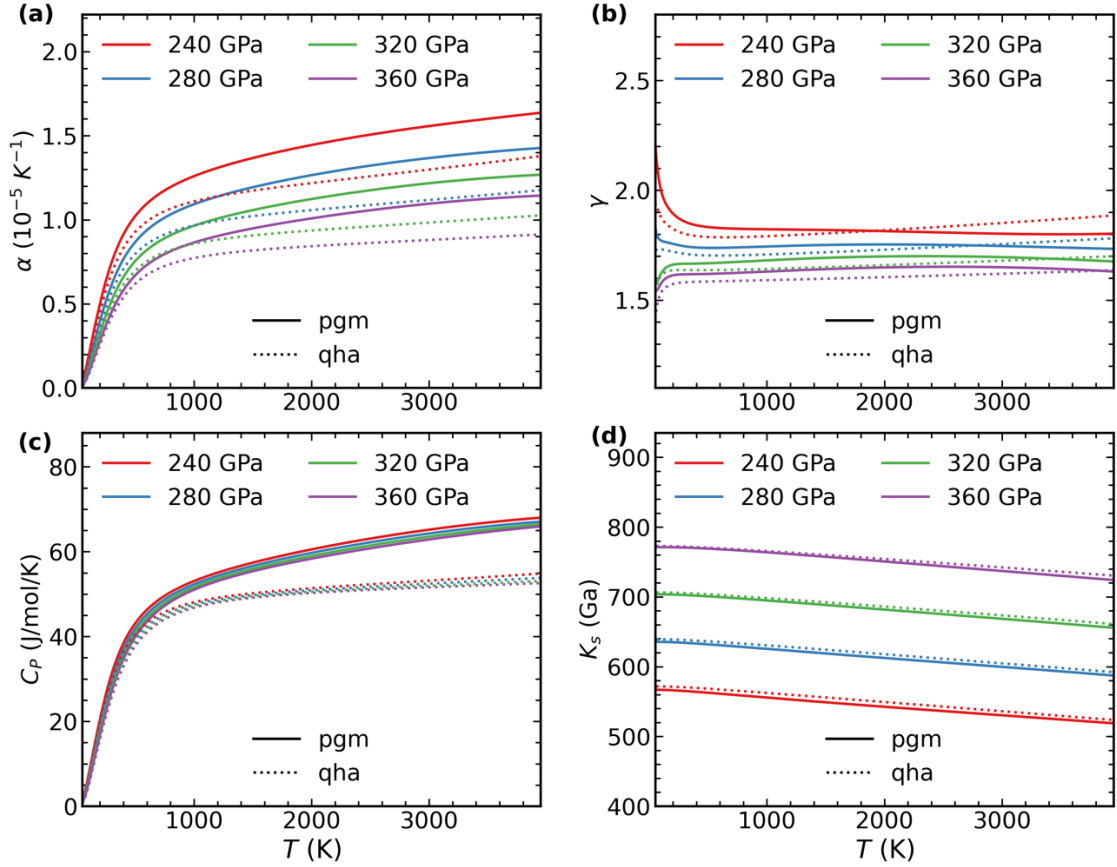

**Figure 9.** Thermodynamic properties of FeO (a) thermal expansivity, $\alpha$, (b) Grüneisen parameter, $\gamma$, (c) isobaric heat capacity, $C_P$, (d) adiabatic bulk modulus, $K_S$, vs. temperatures $T$ at four different pressures (240, 280, 320, 360 GPa), obtained from the 'examples/feo'. All results are obtained using the pgm code. Solid lines include thermal electronic excitation on the free enrgy, $F(V,T) = E_0(V) + E_{zp}(V) - \int_0^T S(V,T')dT' = F_{mermin}(V,T) + F_{vib}(\tilde{\omega}(V,T),T)$. Dotted lines do not include the effect of electronic excitation and are also identically reproduced by the qha code [37], a direct implementation of the quasiharmonic approximation, with $F(V,T) = E(V) + F_{vib}(\omega(V),T)$.